# scientific reports

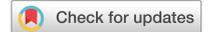

**OPEN**

# The role and contribution of magnetic fields, characterized via their magnetic flux, to the statistical structuring of the solar atmosphere

K. J. Li[1]✉, J. C. Xu[1] & W. Feng[2]

The anomalous heating of the solar upper atmosphere is one of the eight key problems in modern astronomy. Moreover, the stratification of the solar atmosphere is an outstanding key-problem in solar physics. In this study, a hot butterfly-like pattern is found to run through the chromosphere to the corona lying right on top of the magnetic butterfly pattern of sunspots in the photosphere. We thus propose to introduce the term butterfly body to describe the butterfly diagram in the 3-dimensional atmosphere. Besides, we discuss the so-called polar brightening in different layers. It is found to be statistically in anti-phase with the solar cycle in the photosphere and the chromosphere, while in phase with the solar cycle in the corona. Accordingly, we describe the role and relationship of solar magnetic elements of different magnetic flux strengths to explain the statistical structuring of the solar atmosphere with the butterfly body over the solar cycle.

Analyses of full-disk magnetographs indicate that solar magnetic fields could be divided into five categories: one is large-scale magnetic elements in active regions including sunspots (called Component-I hereafter), and the other four are small-scale magnetic elements outside active regions[1,2]. Among the latter four categories, in solar cycle 23 those two whose magnetic flux is respectively in the range of $(4.27 - 38.01) \times 10^{19}$ Mx (called Component-II hereafter) and $(2.9 - 32.0) \times 10^{18}$ Mx (called Component-III hereafter), account for 53.46% and 37.40% of the total magnetic flux of small-scale magnetic elements or 15.65% and 77.19% of the total number of small-scale magnetic elements, correspondingly. Therefore, the vast majority of the magnetic fields on the full solar disk consist of these three components. The characteristics of the three components are displayed in Table 1. The spatial-temporal variation of Component-I and II both presents a butterfly diagram in a solar cycle, and their temporal evolution is in phase with the solar cycle. In contrary, Component-III are distributed all over the solar surface and in anti-phase with the solar cycle at low and middle latitudes of $0°$–$60°$[1-3].

The three components (i.e., Component-I, II, and III) are extracted from the 5 min averaged full-disk magnetograms by the Michelson Doppler Imager on board the Solar and Heliospheric Observatory (MDI/SOHO) with a pixel size of $\sim 1.451$ Mm. One magnetogram is used for each day. The data sets span from 1996 September to 2010 February (3764 days in total), i.e., the entire 23rd solar cycle. Basically, bulks of magnetic concentrations are identified according to the magnetic flux of pixels on the magnetograms after they have been smoothed with a boxcar function and corrected by the heliocentric angle of each pixel. The large bulks with strong magnetic fields, at least 15 Mx cm$^{-2}$ on the edges, are classified as active regions (Component-I). Small bulks with weak magnetic fields in quiet Suns are classified as small-scall elements. These small-scale elements further form four groups according to their flux and relation with the solar cycle. Pixels with heliocentric angles larger than 60 degrees are generally excluded due to the low sensitivity, low spatial resolution, and large noise level in the measurements in these areas. For more details on the data extraction process, please refer to Jin et al.[2].

In general, magnetic activities are an acceptable perspective to explain solar violent activities and slow variations[4-7], and accordingly this study will attempt to do so as well. Now that magnetic fields of different categories present different solar-cycle-phase characteristics in their temporal evolutions and space (latitude) distributions, then traces of these characteristics must be left in solar activities and variations due to magnetic fields acting on them. In other words, solar activities/variations with the same solar-cycle-phase characteristic as that of a category of magnetic fields should be statistically caused by the category. This study will look for traces of that magnetic fields of different categories acting on solar activities and variations, based on these solar-cycle-phase characteristics.

[1]Yunnan Observatories, Chinese Academy of Sciences, Kunming 650011, China. [2]Research Center of Analysis and Measurement, Kunming University of Science and Technology, Kunming 650093, China. ✉email: lkj@ynao.ac.cn

| | Component-I | Component-II | Component-III |
|---|---|---|---|
| Criteria or flux range | $\geq 15$ Mx/cm$^2$ in an area larger $9 \times 9$ pixels | $(4.27 - 38.01) \times 10^{19}$ Mx | $(2.9 - 32.0) \times 10^{18}$ Mx |
| Ratio in terms of flux | 47.17% | 28.24% | 19.76% |
| Ratio in terms of number | Negligible | 15.65% | 77.19% |
| Relation with solar cycle | In-phase | In-phase | Anti-phase |
| Magnetic features | Active regions of sunspots | Mainly network magnetic elements, and ephemeral active regions | |

**Table 1.** Components of active regions and small-scale elements.

In the solar atmosphere, the temperature abnormally increases above the photosphere outwards, from the chromosphere to the corona. This is the so-called puzzle of the anomalous heating of the upper atmosphere (the chromosphere and the corona), one of the eight key mysteries in modern astronomy which has long vexed scientists[6]. For the anomalous heating, modern high-resolution and high-cadence observations indicate that the upper atmosphere should be mainly heated by the ubiquitous small-scale magnetic activity events, such as spicules, nano-flares, petty tornados/cyclones, and so on[8,9] (references therein), which are closely related to small-scale magnetic elements. Strictly speaking, modern observations have just demonstrated that these events may contribute to the heating of the upper atmosphere by providing channels and becoming a class of candidates for the heating; however, they are merely individual local activities, and it has not been illustrated that the upper atmosphere is effectively heated as a whole by them. Actually, the following issues need to be clarified for the anomalous heating of the entire upper atmosphere: why it always (forever) remains hotter than the photosphere; This point involves the temporal evolution of the global upper atmosphere! The way to answer this question might be via analyzing the long-term evolution of the whole upper atmosphere. However, up to now, the long-term evolution of global observations has been rarely investigated to address the issue. Recent statistical studies found that long-term variation of the heated corona keeps step with long-term variation of small-scale magnetic fields, confirming that the corona can be effectively heated by small-scale magnetic activities[8]; furthermore, it is such heating mechanism that causes the corona to rotate faster than the underlying photosphere[9]. However, these statistical studies have ignored details of magnetic fields of various scales/categories on the solar disk, and the effect of large-scale magnetic fields is masked by small-scale magnetic fields.

In observations, the solar atmosphere is divided into the photosphere, the chromosphere, and the corona. The stratification of the solar atmosphere is well known, but so far no explanations have been given to the reason why the atmosphere is stratified in such a way.

Many solar activity phenomena present the solar cycle, appearing in different layers of the atmosphere, and they are in phase or in anti-phase with the solar cycle. In this study, in order to investigate the role of magnetic fields of various categories, we will survey solar-cycle-phase behaviors (in phase or in anti-phase) of various solar activity phenomena, especially including corresponding phenomena of the high-layer atmosphere above the butterfly diagram of sunspots and polar brightening, and they are believed to be caused by magnetic elements of different categories. Next, the solar-cycle-phase behaviors of these phenomena are compared with those of magnetic elements of different categories to determine which category of magnetic elements are connected to which phenomenon, causing it to present its solar-cycle behavior, and meanwhile function of magnetic elements of different categories should be endowed. Accordingly, many important issues including those mentioned above will be addressed.

## The active atmosphere
### Examination of the active atmosphere and proposal of butterfly body.
The first butterfly diagram appeared as the Spörer's law of zones showing latitudinal drift of sunspots, i.e., large-scale magnetic active regions in the photosphere, and it later developed into the famous Maunder's butterfly diagram[10]. Solar eruption events are generally related to magnetic activities of sunspot regions, and they may be called as or belong to the active atmosphere, which statistically refers to the butterfly-pattern area. Area outside the butterfly pattern generally belongs to the quiet atmosphere.

The butterfly diagram in the chromosphere was proposed with the Ca II K full-disk maps of network bright elements observed at the Mount Wilson Observatory[11] and the Kodaikanal Solar Observatory[12]. Network bright elements are several hundreds to nearly one thousand degrees hotter than the background chromosphere. These two independent observations indicate that "butterflies" in the chromosphere, the low layer of the anomalously-heated solar atmosphere, are "hot" structures (relative to the background chromosphere). The full-disk solar chromosphere has also been observed respectively at radio 17 GHz (1.76 cm) of the Nobeyama Radioheliograph at the National Astronomical Observatory of Japan[13] and at radio 37 GHz (0.81 cm) of the 13.7-m Cassegrain radio telescope at the Aalto University Metsähovi Radio Observatory in Finland[14]. The temporal-latitudinal distribution of radio synoptic maps at each of these two frequencies presents a butterfly diagram, and they also reveal that "butterflies" are hot structures in the chromosphere[13–18]. Radio flux at 17 GHz measured from active regions may come from the chromosphere and the corona[13]. The radio synoptic maps are composed from radio maps, on which radio brightenings are displayed as strong features with high brightness temperatures. A solar radio map records the full-disk intensities at a usable frequency of the telescope. The equivalent temperatures can be calculated with the intensities on the map. The measurements are usually scaled relative to the quiet sun level which is the median intensity (temperature) of the entire map, so that the observations are comparable over the years[14].

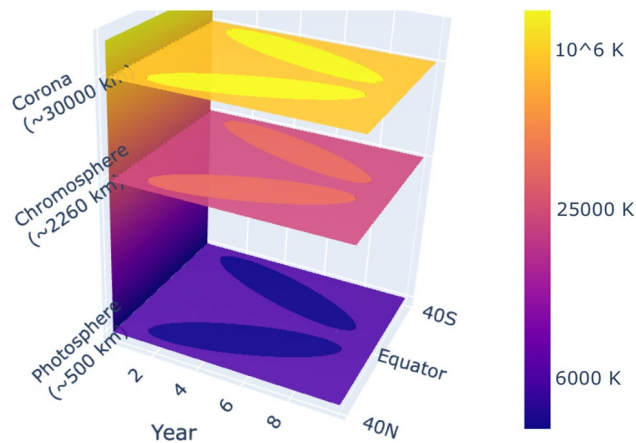

**Figure 1.** Illustration of the 3-dimensional solar butterfly body throughout the solar atmosphere from the photosphere to the corona. The *x*-axis roughly indicates time-unit years in the 11-year solar cycle, the *y*-axis indicates latitudes within 40° (N: the northern hemisphere, and S: the southern), and the *z*-axis indicates height from above the bottom of the photosphere.

The green coronal line (Fe XIV, 5303 Å), which is the best diagnostic tool for describing activities of low corona and representing the irradiance energy emitted through the green corona, has been used to observe the low corona at an altitude of about 30,000 km since the 1940s[19–21]. The observations of coronal green line intensity respectively at Lomnicky Stit Station[22–24], Pic du Midi[19], and Kislovodsk solar station[25], all indicate that the temporal-latitudinal distribution of coronal green line intensity presents a butterfly diagram, and "butterflies" in the low solar corona are "hot" structures (relative to the background corona). The forbidden coronal line Fe X (6374 Å) presents the same phenomenon as these green coronal observations[26].

During cycle 24, the full-disk solar corona has been observed by the Solar Dynamics Observatory/Atmospheric Imaging Assembly at extreme ultra-violet (EUV) bands, such as 193 Å, 171 Å, and so on, and 193 Å and 171 Å are known to have temperature responses peak at $\sim 10^6$ K and $\sim 6 \times 10^4$ K, respectively[17]. The temporal-latitudinal distribution of EUV intensity at each of 193 Å and 171 Å channels presents a butterfly diagram, and "butterflies" are "hot" structures in the corona[17].

To sum up, interestingly, the butterfly diagram that people are used to referring to the photospheric structures of active regions of sunspots is found to actually show a 3-dimensional (3D) butterfly-shaped structure in the solar atmosphere which is illustrated in Fig. 1, which is here called the butterfly body. In the photosphere, the 3D butterfly body shows cold magnetic structures, but in contrast, it presents relatively hot structures in the anomalously-heated solar atmosphere, that is, their temperature is correspondingly higher than the background chromosphere and corona.

In a solar cycle in the anomalously-heated solar atmosphere, the appearance of the heated butterfly body is almost the same as that of the photospheric butterfly diagram, which implies that the energy that is responsible for the anomalous heating of the active atmosphere (the butterfly body) should certainly originate from magnetic activities of the photospheric butterfly diagram. If the energy comes from activities outside the butterfly body, it would be impossible to maintain a consistent body shape (butterfly). Therefore, up to now, this is the most direct and the most obvious global evidence that the heating of the active atmosphere can be caused by magnetic activity in a long run.

Since both Component-I and II present a butterfly pattern, a question then arises: of the two components, which one plays a major role in heating the chromosphere/corona? Next, we will attempt to address this issue from the perspective of magnetic activities.

**Anatomy of butterflies.** Data used in this study include the coronal index (CI) from Jan 1, 1939 to Dec 31, 2008, synoptic magnetographs (denoted as $MF_s$ hereafter) from Carrington rotations 1625 to 2078 (Feb. 1975–Dec. 2008), and the small-scale magnetic elements (Component-II) of $(4.27 - 38.01) \times 10^{19}$ Mx at latitudes −60° to 60° in the period of Sept. 1996–Dec. 2008[1]. The large-scale magnetic fields of sunspots in active regions exist only in $MF_s$, thus here $MF_s$ is utilized to represent to some extent the large-scale magnetic fields in active regions, and synoptic magnetographs come from NSO/kitt peak (for details, please see https://www.nso.edu/data/historical-archive/).

CI was derived from observations of coronal intensity at the green coronal line (5303 Å) by ground-based coronagraphs, and coronal intensity was measured daily at a height of 40″ above the solar limb at 72 evenly spaced points on a circle, which start from 90° (the north pole) and reduce by 5° at a time at the east lime until −90° (the south pole), and then increase by 5° at a time at the west limb until 85°. CI was given in millionths of intensity of the solar disk (coronal units) and then converted to the photometrical scale of Lomnicky Stit Station, and its time series can be downloaded from the web site of the National Geophysical Data Center of NOAA: https://www.ngdc.noaa.gov/stp/space-weather/solar-data/solar-indices/solar_corona/5-degree-data/. CI represents daily irradiance of the green corona at 5303 Å, and it is called energetic index[22,23].

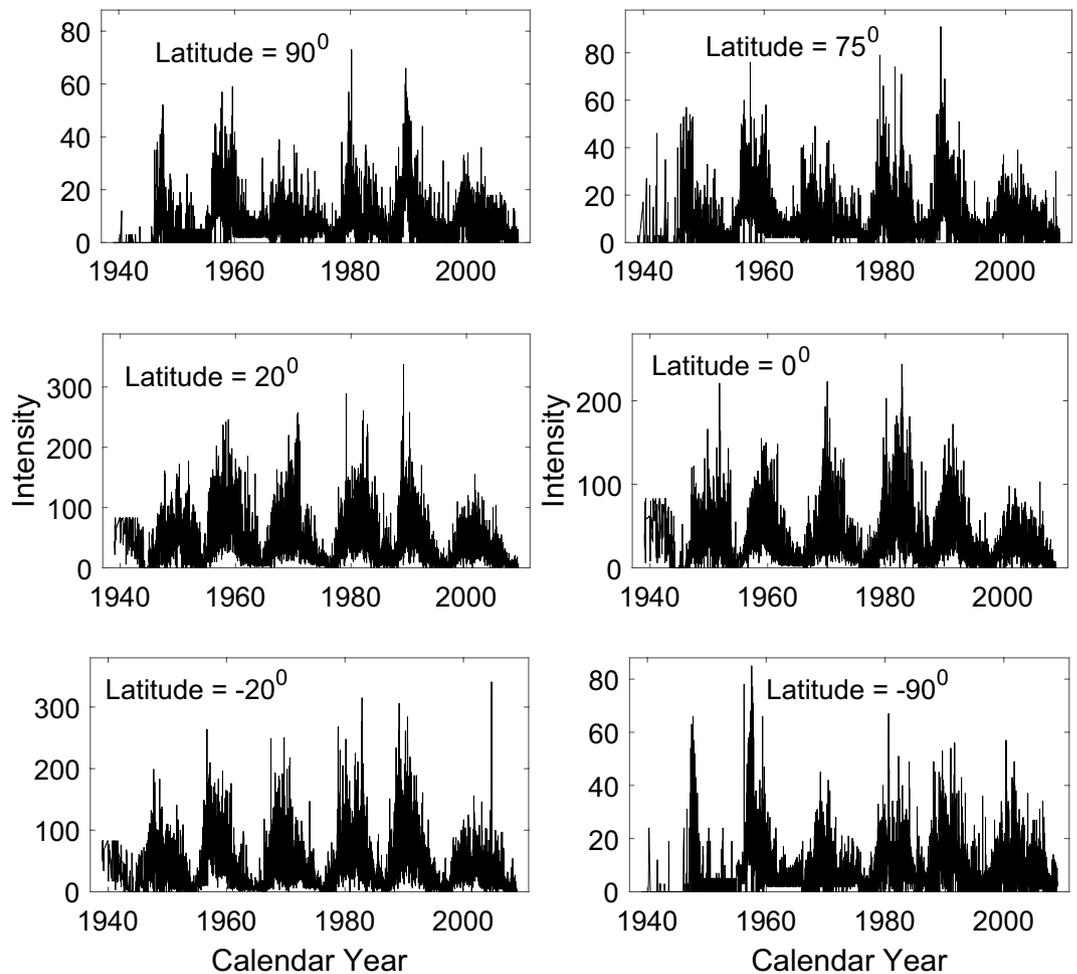

**Figure 2.** Time series of CI at 6 latitudes. The latitudes are shown at the top of each panel.

Figure 2 shows time series of monthly mean CI at 6 latitudes with latitudes in the southern hemisphere given in negative values, and Fig. 3 displays the time-latitude distributions of magnetic fields ($MF_s$ and Component-II) and CI. Generally, CI at low latitudes is obviously (8–10 times) larger than that at high latitudes. These two figures indicate the existence of the solar cycle in the temporal variation and the spatial-temporal distributions of activity indexes. $MF_s$, Component-II, and CI all present a butterfly diagram. Next, their monthly mean values at the 72 latitudes are used to investigate relationship between butterfly diagrams respectively in the photosphere and the corona.

Cross-correlation analysis is made between latitude distribution of CI respectively with that of $MF_s$ and that of Component-II in each month of their common time span (Sept. 1996–Dec. 2008), and Fig. 4 shows the obtained correlation coefficients (CCs). The 95% confidence level is also shown in the figure, and CC is statistically significant at the 95% confidence level in most months of a solar cycle. Among CCs of statistical significance, CC of CI and Component-II ($CC_{II}$) is larger than the corresponding CC of CI and $MF_s$ ($CC_{MF}$) in most months, and just during Jan. 1998–June 1999 $CC_{MF}$ is larger.

In this study, we made use of the sample Pearson correlation coefficient which measures the linear correlation between two data sets. For example, the correlation coefficient $r_{xy}$ between two given data sets $x$ and $y$ with the same length, namely $\{x_1, x_2, \ldots, x_n\}$ and $\{y_1, y_2, \ldots, y_n\}$, is defined as $r_{xy} = \frac{\sum_{i=1}^{n}(x_i - \bar{x})(y_i - \bar{y})}{n\sigma_x\sigma_y}$. In the formula, $n$ is the number of elements in each of the data set, and $\bar{x}$ and $\sigma_x$ denote the mean value and standard deviation of data set $x$ (the meanings of $\bar{y}$ and $\sigma_y$ are similar).

The Fisher translation[27] is used to test statistical significance for the difference of paired CCs ($CC_{II}$ vs. $CC_{MF}$) at 95% probability. For CI measured at the east solar limb, $CC_{II}$ is statistically significantly larger than $CC_{MF}$ in 6 months among the total 156 months, although most of $CC_{II}$s are numerically larger than the corresponding $CC_{MF}$. For CI measured at the west solar limb, $CC_{II}$ is significantly larger in 9 months, although most $CC_{II}$s are larger in number. No $CC_{MF}$s are significantly larger. Therefore in general, CI is more closely related to Component-II than to $MF_s$ in all probability.

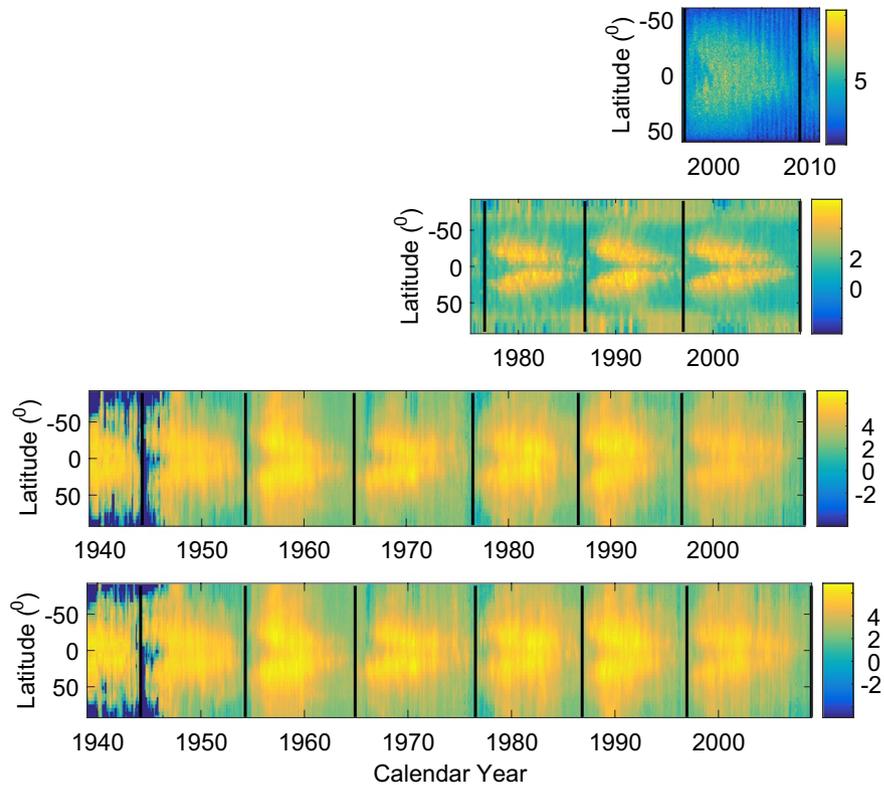

**Figure 3.** Time-latitude distribution of solar measurement quantities of four kinds. The first (top) panel: the small-scale magnetic elements (Component-II) whose fluxes span $(4.27 - 38.01) \times 10^{19}$ Mx. The second: the solar magnetic field ($MF_s$) from synoptic magnetographs. The third: CI measured at the east limb. And the fourth (bottom) panel: CI at the west limb. The vertical thick dashed lines show the minimum times of the solar cycles.

Correlation analysis is also carried out between time series of CI measured at the east solar limb in the common time span respectively with that of $MF_s$ and that of Component-II, at a certain latitude, and then similarly, correlation analysis is carried out with CI measured at the east solar limb replaced by CI at the west limb. During the common time span at a latitude, $MF_s$ is always available, but Component-II is not measured in 7 months. When Component-II is unavailable in a month, $MF_s$ is assumed to have no data in the month, so that the same amount of data are always used to calculate $CC$ at a latitude. Figure 5 shows the obtained $CC$s and their 95% confidence level line. For latitudes not higher than 40°, $CC$ of CI and Component-II is larger than the corresponding $CC$ of CI and $MF_s$ in 13 cases of the total 17. The Fisher translation is also used to test statistical significance at 95% probability for the difference of paired $CC$s ($CC_{II}$ vs. $CC_{MF}$) in this case. For CI measured at the east solar limb, $CC_{II}$ is statistically significantly larger than $CC_{MF}$ at 5 latitudes, and for CI measured at the west solar limb, $CC_{II}$ is significantly larger at 7 latitudes. No $CC_{MF}$s are significantly larger. Therefore generally, CI seems more closely related to Component-II than to $MF_s$ at low latitudes, where sunspots appear.

To sum up, on the monthly scale, the butterfly diagram of CI is more related to the magnetic fields of the Component-II than to the magnetic fields of sunspots, but contribution of the latter to the formation of the CI butterfly diagram cannot be excluded. This is especially true on the scale of days and/or smaller, because there are no sunspots on the solar disk in some days.

## The quiet atmosphere
### Survey of the quiet atmosphere.
The He I 10,830 Å line was utilized to routinely observe the solar chromosphere at National Solar Observatory located on the Kitt Peak of Arizona, and a series of synoptic maps of He I 10,830 Å intensity were obtained. Through analyzing the maps, Li and Feng[28] found that He I intensity presents a butterfly pattern, also indicating that hot butterfly structures appear rightly above the butterfly pattern of sunspots, and the long-term variation of the active chromosphere in general sense, is in phase with the solar cycle. Furthermore, the long-term variation of the quiet full-disk chromosphere described by He I intensity was found to be also in phase with the solar cycle[28]. Because the He I line is an absorption line on the disk of the chromosphere, the long-term variation of the generally referred quiet chromosphere (the quiet chromosphere observed in $H_\alpha$, Ca II K or other emission lines) is in anti-phase with the solar cycle. Thus a different kind of magnetic field class should exist and be able to explain such anti-phase correlation. Indeed we can find the solu-tion in the magnetic elements of Component III which are measured to be in anti-phase with the solar cycle and distributed all over the solar disk[1,2]. Moreover, it is inferred that they are responsible for the abnormal heating of the quiet chromosphere[28].

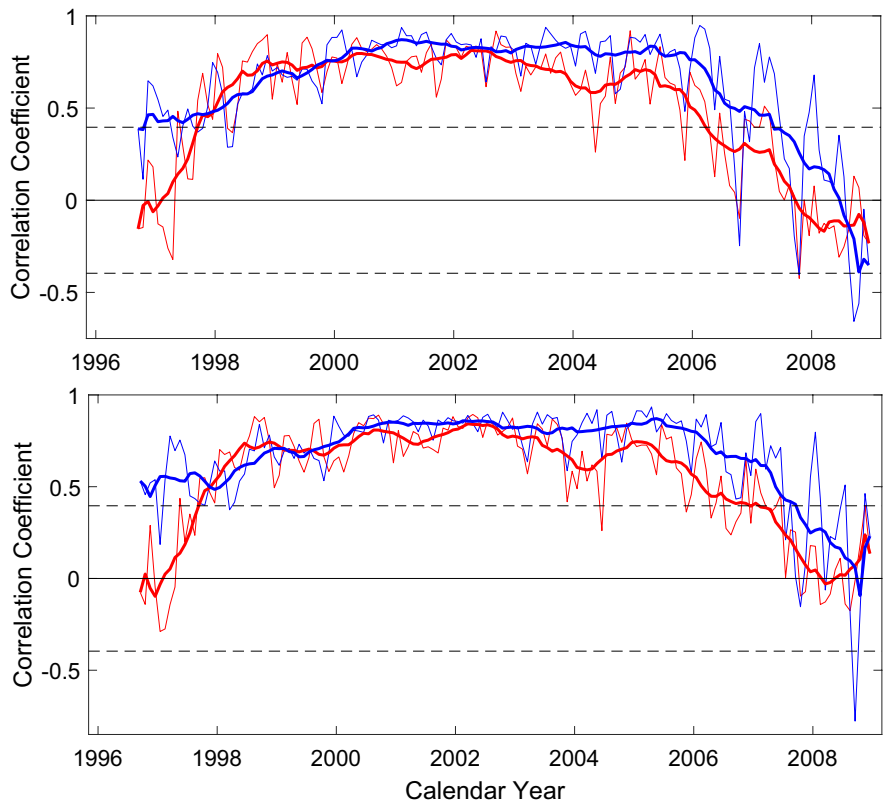

**Figure 4.** Top panel: cross correlation coefficient between latitude distribution of CI measured at the east solar limb, respectively with the magnetic field $MF_s$ (red thin line) and the small-scale magnetic elements (Component-II, blue thin line), at a certain month. The thick red/blue line is the 12-point smoothing of the corresponding thin red/blue line, and the two horizontal dashed lines show the 95% confidence level. Bottom panel: the same as the top panel, but CI measured at the east solar limb is replaced by CI at the western limb.

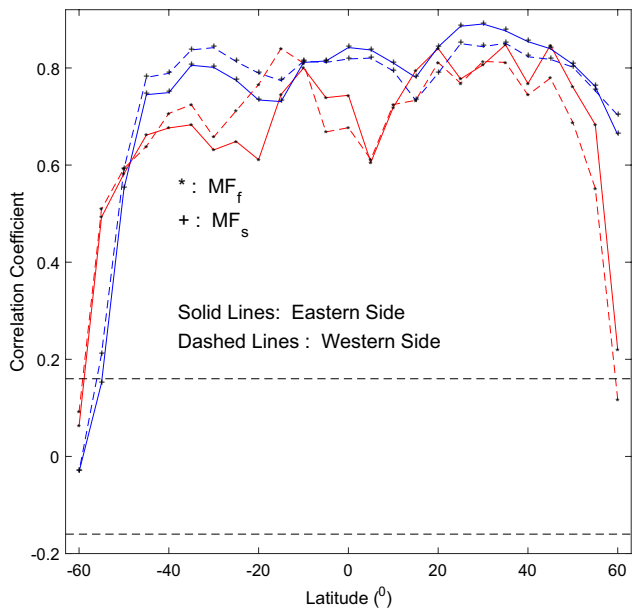

**Figure 5.** Cross correlation coefficient between time series of CI measured at the east/west solar limbs (solid/dashed lines), respectively with the magnetic field $MF_s$ (asterisks in red lines) and the small-scale magnetic elements Component-II (plus signs in blue lines), at a certain latitude. The two horizontal dashed lines show the 95% confidence level.

Recently, Li et al.[29] found that the quiet chromosphere abnormally rotates faster than both the photospheric plasma as well as its magnetic elements. Moreover, they deduced that the small-scale magnetic elements of Component-III rooted in the leptocline (the layer between the solar photosphere and the upper convective zone, please see Godier and Rozelot[30]) might be responsible for that faster rotation. Accordingly their final conclusion has been that the quiet chromosphere should be heated mainly by the magnetic elements of component III.

Li et al.[8] found that the heated corona statistically synchronizes with small-scale magnetic elements which are mainly from Component-II, possibly plus Component-III, indicating that the corona is heated mainly by small-scale magnetic elements. Li et al.[9] found the coronal rotation is faster than the underlying photosphere, and consequently they inferred that the corona is heated mainly by small-scale magnetic elements. Contemporary high-resolution observations clearly show that the corona is being heated by small-scale magnetic activities related to magnetic reconnection and MHD wave processes[8,9,31] (references therein). Therefore, observational, statistical, and theoretical studies all hint to the relationship between the abnormal heating of the quiet corona to small-scale magnetic activities.

Utilizing intensities of five coronal spectral lines of about 400 days during the minimum time of the solar cycle 23, Mancuso et al.[32] found that the quiet corona rotates faster than the photospheric plasma, which suggests that the quiet corona is mainly heated by small-scale magnetic structures rooted near $0.99R_\odot$ (also in the leptocline). During the minimum of a solar cycle, the solar disk is mainly populated by small-scale magnetic elements of Component-III[1,2]. Thus their finding[32] actually implies that the quiet corona should be heated mainly by the elements of Component-III during the minimum time of a solar cycle.

As shown in Fig. 4, around the minimum time of a solar cycle, latitudinal distribution of CI in a month is unrelated or even negatively correlated with that of Component-I and even with that of Component-II. Moreover, the solar disk is populated mainly with small-scale magnetic elements of Component-III (see the Fig. 5 in Jin et al.[1]). In Fig. 4 of Jin and Wang[2], the authors depict the latitudinal behaviour of this Component III during the minimum time of the solar cycle featuring opposing trends when compared to the distribution of sunspots (Component I). Given those facts, it is believed that the latitudinal distribution of CI follows more closely the trend of Component III yielding our belief that during the minimum time of a solar cycle the atmosphere should be heated mainly by Component III magnetic fields.

**Polar brightenings.** Polar brightening is a special phenomenon in the solar quiet atmosphere. It appears in the chromosphere at the minimum time of a solar cycle manifesting itself in observations in the Ca II K line[11,12] and the radio 17 GHz[13,15–18] band. Thus, the observed long-term variation of chromospheric brightness is in anti-phase with the solar cycle around the solar poles. In addition, this finding is supported by polar facula counts (chromospheric flocculi are the upward extension of photospheric faculae[32]) featuring a similar anti-phase behavior. Polar brightening appears in the photosphere also in the minimum time of a solar cycle.

However, in sharp contrast, observations taken in the corona, within the green coronal line (Fe XIV, 5303 Å), the forbidden coronal line (Fe X, 6374 Å), and the EUV 193 Å and 171 Å lines, respectively, all show that polar brightening takes place in the corona in the maximum time of a solar cycle, indicating that long-term variation of coronal brightness is in phase with the solar cycle around the poles[19–26]. Polar brightening is actually manifested as "polar layered temporally-staggered brightening".

For different latitudinal distributions of a solar activity index, the transition from a positive or negative correlation with the solar cycle to its opposite correlation usually goes through a particular band of latitudes, where statistically insignificant correlation appears. For example, polar facula counts evolve over time in such a way[33]. Statistically insignificant correlation does not appear yet from low latitudes till the latitude of 60° for Component-II and III[1,2]. Therefore around the poles, the small-scale magnetic elements of Component-II are believed to be still in phase with the solar cycle, and the small-scale magnetic elements of Component-III are inferred to be still in anti-phase with the solar cycle, which is somewhat in agreement with the quiet chromosphere observation in the He I line[28]. Synoptic magnetographs show that polar fields are in anti-phase with the solar cycle, and they are believed to be made up mainly of Component-III. While the polar brightening seen in the corona should be, in all probability, attributed mainly to Component-II, the one observed in the chromosphere should be attributed mainly to Component-III, based on their same phase relation with the solar cycle. In order to discover long-term evolution of solar activities and the magnetic fields at the poles, long-term observations of solar polar regions from the polar orbit of the Sun are indispensable in the future.

When the small-scale magnetic elements of Component-III reach the maximum at the minimum time of a solar cycle, the averaged CI over the full solar disk is just about 10% of the CI averaged over the full solar disk at the maximum time of the solar cycle[20,21], and meanwhile around the solar poles, monthly mean CI is just about 10% of the monthly mean CI at the maximum time (see Fig. 2). Even if it is assumed that CI at the minimum time of a solar cycle is contributed completely by these small-scale magnetic elements (they peak in this time), their contribution to the coronal heating is still very small during a solar cycle. Therefore the contribution to the coronal heating by these magnetic elements (Component-III) is believed to be very small in a solar cycle. The magnetic elements of Component-II peak at the maximum time of a solar cycle, however there is no significant brightening around the poles of the chromosphere. Therefore the contribution to the quiet chromosphere heated by the magnetic elements of Component-II is inferred to be very small during a solar cycle. This indicates that, the greater the flux of magnetic fields of a category is, the higher the altitude in the atmosphere is, where they will create an impact and lead to local heating.

## Conclusions and discussion

Sunspots constitute the first historically observed phenomenon among solar activities; long-term observations of sunspots lead to the birth of the butterfly diagram of sunspots, and it is the first form of butterfly diagram in the solar atmosphere. This study investigates the spatial and temporal distribution of solar activities in the chromosphere and the corona, where the solar atmosphere is being anomalously heated.

In the chromosphere, long-term full-disk observations respectively in the Ca II K line, radio 17 GHz, and radio 34 GHz all show that a butterfly pattern also exists but presents structures hotter than their chromosphere background, which are in sharp contrast to the cold structures in the photosphere. The full-disk maps observed respectively in the green coronal line (5303 Å), the forbidden coronal line (Fe X 6374 Å), EUV 193 Å, and EUV 171 Å all indicate the appearance of a butterfly pattern in the corona, and its butterfly structures are hotter than the coronal background. Semi-empirical models of sunspot penumbra and umbra also indicate that temperature of sunspot penumbra and umbra is higher than their background atmosphere in the upper chromosphere and the transition region[34]. Therefore a three-dimensional butterfly body is proposed to form in the whole solar atmosphere, from the photosphere to the corona, and the butterfly diagram of sunspots in the photosphere is its pedestal. It is interesting that what lays right on top of a magnetic butterfly is precisely a hot butterfly body. The energy that heats the butterfly body in the abnormally-heated atmosphere must come from magnetic activities of the photospheric butterfly of sunspots, and this is the most direct and the most evident global observation evidence that the active atmosphere is being heated by solar magnetic activities (Component-II plus Component-I).

In this study, CI is found to be seemingly more closely correlated with the small-scale magnetic elements of Component-II than with the large-scale magnetic field of sunspots, therefore the active corona may probably be mainly heated by the small-scale magnetic elements of Component-II. Sunspots may decay into small-scale magnetic elements, therefore statistically, the correlation coefficient of the heated atmosphere with strong magnetic fields becomes larger after a solar rotation than that when the sunspots are first observed[8], supporting the heating of the active corona mainly by Component-II. At present, there is no evidence that the active chromosphere is mainly heated by component-II as the active corona does, and the heating of the active atmosphere is an open question.

Rotation characteristics of the He I chromosphere[29] and its phase relationship with the solar activity cycle[28] indicate that the quiet chromosphere is heated mainly by the small-scale magnetic elements of Component-III, which is somewhat confirmed by the polar brightening in the chromosphere. In the chromosphere and the photosphere the polar brightening is in anti-phase with the solar cycle, but in the corona the polar brightening is in phase with the solar cycle, showing "polar layered temporally-staggered brightening". The brightening in the solar atmosphere indicates that the small-scale magnetic elements of Component-II mainly heat the corona. Contrarily, their contribution to the chromospheric heating is relatively small, as small-scale magnetic field elements of Component-III are mainly responsible for that chromospheric heating. We note here that the contribution of component III fields to the coronal heating is negligible. This standpoint is also confirmed by the result coming from the correlation analysis between CI and solar magnetic fields: Component-II is the key contributor in the remaining time of a solar cycle other than its minimum time, and Component-III is the major contributor just in the minimum time. Multi-wavelength observations found that the coronal heating by reconnections of small-scale magnetic fields could lead to the chromospheric heating[35] (references therein), and this is the observational evidence of the standpoint.

Based on the standpoint, we come to the conclusion that the stratification of the solar atmosphere is due to magnetic activities of different scales. The photosphere formation is the natural result of the black-body radiation from the Sun as a star of about 5800°. The small-scale elements of Component-III emerge to heat the solar atmosphere, which leads to the formation of the quiet chromosphere, and their heating height determines the formation height of the quiet chromosphere (as mentioned above, their contribution to the coronal heating is very small). The number ratio of these magnetic elements account for a large proportion (77.2%) of all small-scale magnetic elements on the disk in a solar cycle[1], and this is conducive to the formation of the quiet chromosphere. The small-scale elements of Component-II and active regions of sunspots emerge to heat the atmosphere more vigorously, due to their much larger magnetic fluxes than Component-III, which leads to the formation of the active atmosphere from the photosphere, through the chromosphere, to the corona, and thus a butterfly body appears in a solar cycle. Long-term evolution of the solar active atmosphere is in phase with the solar cycle at all atmospheric layers. It is interesting to note that in the chromosphere, there are two totally opposite solar-cycle-evolution phases: the active chromospheric being in-phase with the solar cycle and the quiet chromosphere being anti-phase. This is inferred to be the reason why some solar activities/indexes are positively related with the solar cycle, some are negatively correlated, and some show no cyclic variations. The anomalous heating of the solar atmosphere is essentially the same issue as its stratification, that is, the anomalous heating of the solar atmosphere by different magnetic activities leads to its stratification. The strength (flux size) of magnetic elements determines the height and the degree (temperature) of their acting (heating); and the stronger the strength is, the higher the acting can reach, and the stronger the acting degree is. Such a physical picture of flux size, acting height, and acting degree of magnetic elements is consistent with the observation model of magnetic structures[36,37]. Of course, such a stratification presents a balance symbiosis under interactions of adjacent layers.

As for the formation of transition region lying between the chromosphere and the corona, two inferences and plausible arguments can be given here. First, the flux size of Component-III is an order of magnitude smaller than Component-II, the heating height that Component-III can reach is thus statistically obviously lower than that of Component-I and II, and resultantly a drop/fall, namely the transition region is created. Secondly, there is the fourth component of the magnetic fields on the solar disk actually, whose flux size is $(3.20 - 4.27) \times 10^{19}$ Mx, locating between the size of Component-II and III, therefore the heating height which it can reach is also between Component-II and III. The flux ratio of the fourth to Component-III in a solar cycle is less than one quarter, and the ratio of the forth to Component-II is about one–sixth[1], therefore the contribution by the fourth component to heating the atmosphere is much smaller than that by Component-III and that by Component-II also. The total flux of the Component-I and II in a solar cycle is about 4 times larger than that of Component-III[1], and thus the coronal heating by Component-I and II is much more intense than the heating of the quiet chromosphere by Component-III. This is inferred to be the reason why the temperature suddenly and rapidly increases from the chromosphere to the corona. Due to this strong temperature increase in such a thin layer of the Sun—the transition region—that thin layer is highly dynamic, and it should be possible for the energy to be transported down from the corona to the chromosphere via convective motion or conduction It may be the reason why distinct and stable red-shifts are observed in the transition region[38,39].

| | Component-I | Component-II | Component-III |
|---|---|---|---|
| Active chromosphere and corona | Large | Large | Very small |
| Quiet chromosphere | – | Very small | Very large |
| Quiet corona | – | Possibly large | Possibly small |

**Table 2.** Contribution of various magnetic components to the heating of the solar atmosphere.

The contrast of the rotation period of coronal plasma with that of the solar atmosphere at the bottom of the photosphere supports this standpoint[9]. This view is also supported by the contrast of time series of solar spectral irradiances at the spectral intervals 10–390 Å and 1160–24160 Å respectively with time series of large- and small-scale magnetic activities[8]. The chromospheric butterfly pattern of the Ca II K line can extend towards latitudes larger than 55°, showing the so-called extended solar cycles[12]; the same happens to the coronal butterfly pattern of the green coronal line (see Fig. 5 of Li et al.[40]) and also the butterfly pattern of Component-III[40]. Yet this is not the case for the butterfly pattern of sunspots (if small-scale elements are not considered). Therefore, the active solar atmosphere is believed to be mainly heated by small-scale magnetic activity, from the perspective of the evolution of the global solar activity mentioned above. That is, of the two kinds of the magnetic butterfly diagrams, the butterfly diagram of the small magnetic elements of Component-II is a major contributor the heating the hot butterfly body above itself.

Recent high-resolution observations of small-scale solar activity events by Transition Region and Coronal Explorer, Solar Dynamics Observatory, Solar Terrestrial Relations Observatory, Hinode, New Vacuum Solar Telescope, and so on have illustrated how the corona is heated[8,9,41] (references therein). On the one hand, solar activities are closely related to magnetic configurations which generally follow self-similar patterns. On the other hand, it is well known that with the decrease of scales, the number of activity events increases exponentially. Both facts considered together are indicative for the importance of small-scale solar activity events in the coronal heating process[42]. Therefore, current high-resolution observational data support the results of this study by means of case analyses. These observations have also been explained theoretically[7].

The sun is magnetic, and its magnetic field elements can be classified into 5 groups according to their magnetic flux content. The main categories are magnetic elements of Component-I, II, and III. These three categories account for more than 95% of the total magnetic flux in a solar cycle[1]. The functions of various magnetic components are summarized as follows and also in Table 2. Statistically, violent eruptions (flares, coronal mass ejections, and so on) are mainly related to activities of Component-I. The active atmosphere shaped by the butterfly body is heated mainly by activities of Component-II plus Component-I, and the effect of Component-II is probably more obvious. The heating and the formation of the quiet chromosphere are mainly attributed to activities of Component-III. The heating and the formation of the quiet corona are mainly attributed to activities of Component-II, but component III seems to contribute a little, and the contribution is obvious, especially in the minimum time of a solar cycle.

This study gives a plausible explanation for the abnormal heating and the stratification of the atmosphere, but there are some conjectures in the explanation that need to be confirmed in the future. In view of this, work to be carried out in the future includes: (1) long-term observations of the magnetic fields and total irradiance around the solar poles; (2) study on long-term evolution of the (background) photosphere; and (3) study on relation of magnetic elements and heating.

## Further discussion and speculation

The solar magnetic fields become more and more horizontal as their height increases from the photosphere to the upper atmosphere, forming the so-called canopies[43,44]. Temperature and density sharply change from the chromosphere to the transition region and the corona, and magnetic canopies separate them on the whole. Therefore the significant differences in physical states for the chromosphere and the corona may determine their different heated methods/modes. The magnetic elements of Component-III are obviously smaller in flux but much more in the number than those of Component-II and those of Component-I as well. Magnetic canopy structures significantly increase possibility for contact between adjacent magnetic fields, compared with their magnetic fields in the lower atmosphere. Thermal gas pressure is obviously larger than magnetic pressure at the bottom of the photosphere but much lower in the corona[43], therefore magnetic lines can easily transfer disturbance moves from the photosphere to the corona. As a result, adjacent magnetic canopies may probably contact each other, and then reconnection happens. Therefore statistically, the quiet chromosphere may be pulsatively heated in a more frequent but less violent way than the corona, and the predominant mechanism for the coronal heating is inferred as magnetic reconnections, but for the quiet chromosphere, magnetohydrodynamic waves are conjectured as a significant factor. Such a scene is supported by observation and theoretical researches[35,45] (references therein). Solar wind is found to originate in coronal magnetic funnels[46], also supporting the scene. Lemaire et al.[47] believed that plage should be additionally heated by dissipation of waves, when they constructed a semi-empirical atmospheric model of a plage, further validating the scene. Hydrogen is the main component in the solar atmosphere, and it is ionized in the upper chromosphere. Ionized particles can hardly cross canopy-shaped magnetic lines[43,44]. Therefore, canopies act somewhat as a "dam" to prevent leakage/diffusion of charged particles from the chromosphere into the corona. This is inferred to be one of the reasons why density sharply decreases from the chromosphere to the transition region and the corona.

The different solar-cycle behaviors of different categories of magnetic elements imply that their origins may be different. At the solar shallow subsurface (the leptocline), the temporal variation of the solar seismic radius is in anti-phase with the solar cycle. The leptocline layer which connects the convective zone with the solar atmosphere, is believed to be the seat of in-situ magnetic fields[48]. Here, the small-scale magnetic elements of Component-III are inferred to originate in the leptocline layer, according to their same phase relationship with the solar cycle, and the leptocline and the tachocline may be two source regions where the creation of the solar magnetic fields is occuring. A strong change in the energy transmission mode, namely from radiative energy transport in the radiative zone to convective energy transport in the convective zone, is occurring in the tachocline, and it is the cradle of sunspots, which are in phase of the solar cycle. It is interesting for the immediately following reversal. A strong change in the energy transmission mode, namely from convective energy transport in the convective zone to radiative transport in the photosphere, is occurring in the leptocline, and it is the cradle of Component-III, which is in anti-phase of the solar cycle.

Interestingly, the solar differential rotation is characterized by the butterfly pattern, and this phenomenon is called the torsional oscillation[40,49]. Its migration track on the disk (latitudinal-temporal distribution) is the same as that of Component-I plus Component-II. We do not know (causality) relation between the two, maybe the two are only projections of a potent phenomenon in the solar interior[40].

$\gamma$ rays are the highest-energy emission from the sun, which are closely related with cosmic rays. $\gamma$ rays observed at the solar disk are found to show two special features: the long-term variation of the $\gamma$-ray flux is in anti-phase with the solar cycle; and the bright, hard-spectrum emission measured in the minimum time of a solar cycle mainly occurs at the equatorial plane of the sun[50]. Considering that the main reason for variability of $\gamma$ rays is magnetic fields in the solar atmosphere[50], it is inferred here that the features are attributed to the magnetic elements of Component-III, because the elements are also in anti-phase with the solar cycle and mainly distributed at low latitudes around the equatorial plane. Seckel et al.[51] thought that cosmic rays should be mirrored by magnetic fields at the right depths to produce the observed $\gamma$ rays, and if the reflection height is too high, there is not enough material to interact with cosmic rays, and if the reflection height is too low, $\gamma$ rays should be absorbed by the sun. Here "the right depth" is inferred to be locate in the transition region above the chromosphere, because the outward decrease of atmospheric density presents a steep fall from the chromosphere to the transition region, where canopies of small-scale magnetic fields like a plane mirror covering the chromosphere reflect incoming $\gamma$ rays.

The chromosphere is found to be flattened equatorially with the prolateness of order of $10^{-3}$[52]. As aforementioned, the formation of the chromosphere is related mainly with Component III, and coincidentally, its magnetic fields evidently decrease from the equator to the poles[2]. Here, the prolate chromosphere is speculated to be linked closely with Component III, and the heating by magnetic activities causes atmospheric expansion. Long-term variation of the solar radius is in anti-phase with the solar cycle, and the relative variation ratio within a solar cycle is of order of $10^{-4}$[53]. Long-term variation of the helioseismic radius in the leptocline is in anti-phase with the solar cycle, and the relative variation ratio within a solar cycle is of order of $10^{-5}$[44]. Here, long-term variation of the solar radius is speculated to be linked closely with Component III, which is also in anti-phase, and the heating by magnetic activities causes atmospheric expansion at the top of the photosphere.

Magnetic fields are known to cause the Sun to violently erupt at different scales, to lose material in various forms of wind[46] (references therein), to rotate in various ways of accelerating[9,32] (references therein), and even to be shaped itself. Magnetic fields also cause the Sun to outward radiate energy in the form of spatial and temporal variations, to be highly structured, to resist invasion of cosmic rays, and so on. The nature of the sun is magnetic, and magnetic fields are very important in setting many behaviors of the sun.

## Data availability

The datasets analysed during the current study are available: (1) synoptic magnetographs can be publicly downloaded from the web site of NSO/kitt peak: https://www.nso.edu/data/; (2) coronal index can be publicly downloaded from the web site of the National Geophysical Data Center of NOAA: https://www.ngdc.noaa.gov/stp/space-weather/solar-data/solar-indices/solar_corona/coronal-index/; and (3) the small-scale magnetic elements (Component-II) come from Dr. Chun-Lan Jin, and request for the data should be addressed to cljin@nao.cas.cn.

nature portfolio

## Acknowledgements

We thank the two anonymous referees for careful reading of the manuscript and constructive comments which improved the original version of the manuscript. The data of the coronal index and the full-disk synoptic magnetographs are courtesy to be downloaded from web sites. The authors would like to express their deep thanks to the staffs of these web sites. We thank Prof. J. X. Wang for helpful discussion and Dr. Chun-Lan Jin for kindly providing the data of small-scale magnetic elements. This work is supported by the National Natural Science Foundation of China (11973085, 11903077, 11803086 and 12273087), Yunnan Fundamental Research Projects (202201AS070042), the Yunnan Ten-Thousand Talents Plan (the Yunling-Scholar Project), the project of the Group for Innovation of Yunnan Province grant 2022HC020, the national project for large scale scientific facilities (2019YFA0405001), and the Chinese Academy of Sciences.

## Author contributions

All authors analyzed the results and K.J.L. wrote the main manuscript text. J.C.X. prepared Fig. 1. All authors reviewed the manuscript.

## Competing interests

The authors declare no competing interests.

## Additional information

**Correspondence** and requests for materials should be addressed to K.J.L.

**Reprints and permissions information** is available at www.nature.com/reprints.

**Publisher's note** Springer Nature remains neutral with regard to jurisdictional claims in published maps and institutional affiliations.